\documentclass{optica-article}

\journal{opticajournal} 

\articletype{Research Article}

\usepackage{lineno}
\linenumbers 
\usepackage{graphicx}%
\usepackage{xcolor}%
\usepackage{textcomp}%
\usepackage{manyfoot}%
\usepackage{booktabs}%
\usepackage{algorithm}%
\usepackage{algorithmicx}%
\usepackage{algpseudocode}%
\usepackage{listings}%
\usepackage{epsfig}

\begin{document}

\nolinenumbers
\title{Non-Markovian environment induced Schr\"{o}dinger cat state transfer
in an optical Newton's cradle}

\author{Xinyu Zhao and Yan Xia\authormark{*}}
\address{Fujian Key Laboratory of Quantum Information and Quantum Optics (Fuzhou University), Fuzhou 350116, China\\
Department of Physics, Fuzhou University, Fuzhou 350116, China\\}
\email{\authormark{*}xia-208@163.com} 


\begin{abstract*} 
In this manuscript, we study the Schr\"{o}dinger cat state transfer
in a quantum optical version of Newton's cradle in non-Markovian environment.
Based on a non-Markovian master equation, we show that the cat state
can be transferred purely through the memory effect of the non-Markovian
common environment, even without any direct couplings between neighbor
cavities. The mechanism of the environment induced cat state transfer
is analyzed both analytically and numerically to demonstrate that
the transfer is a unique phenomenon in non-Markovian regime. From
this example, the non-Markovian environment is shown to be qualitatively
different from the Markovian environment reflected by the finite versus
zero residue coherence. Besides, we also show the influence of environmental
parameters are crucial for the transfer. We hope the cat state transfer
studied in this work may shed more light on the fundamental difference
between non-Markovian and Markovian environments.

\end{abstract*}

\section{\label{sec:Intorduction}Introduction}

Schr\"{o}dinger cat state \cite{Schroedinger1935N,Monroe1996S} is
a fundamental topic in quantum mechanics since it may reveal the boundary
between the quantum and the classical realms as well as the transition
from one to the other \cite{Zurek1991PT,Zurek2003RMP,Schlosshauer2019,Raimond1997,Haroche2002,Raimond2001}.
As a result, cat state has attracted a lot of research interest \cite{Pezze2019PRL,Zhang2021QSaT,Wang2022SA,Sun2021PRL,Chen2021PRL,Chen2021PRR,Haroche2008,Assemat2019PRL}, various theoretical and experimental studies are made to observe such a superposition state.
So far, a lot of analogs of Schr\"{o}dinger cat states have been
demonstrated experimentally in optical system \cite{Du1997OC,Fu2002JoMO,Grimm2020N,Paris1999JoOBaSO,Zapletal2022PQ,Zhang2021QSaT},
atomic system \cite{Pezze2019PRL}, magnon system \cite{Sun2021PRL,Xu2023PRL}, solid state system \cite{Yang2024NP,Cosacchi2021PRR,Bild2023S},
and many other systems \cite{Cosacchi2021PRR,RomeroIsart2010NJoP}, even including bio-organic systems \cite{RomeroIsart2010NJoP}.

The investigations on Schr\"{o}dinger cat state are not only crucial
for testing fundamental concepts of quantum mechanics, but also valuable
for applications in quantum computing \cite{Darmawan2021PQ,Du1997OC,Fu2002JoMO,Goto2019PRA,Grimm2020N,Kewming2020NJoP,Paris1999JoOBaSO,Puri2017NQI,Xu2022PRR,Zapletal2022PQ,Chen2024PRL,Zhou2022PRA,Qin2021PRL,Zhou2021PRA,Kang2022PRR,Ku2020npjQI,Wang2017PRB}.
For example, cat states are utilized for encoding quantum information
and correcting quantum errors \cite{Darmawan2021PQ,Grimm2020N,Kewming2020NJoP,Grimsmo2020PRX}.
Fault tolerant quantum computation based on cat state qubit \cite{Lund2008PRL,Mirrahimi2014NJoP,Grimm2020N}
shows huge potential of beating the break-even point of quantum error
correction \cite{Ni2023N}. At the application level, besides the
cat state preparation \cite{Chao2019AP,Xiong2019OE,Yang2019JoSA}, the transfer of Schr\"{o}dinger cat state
could be equally important \cite{He2020QIP,Jeong2006PRL,Liu2020FoP,Ran2016SR}.
Because in a quantum processor, the information has to be transported
from one processing unit to another frequently, so the cat state as
the information carrier must be faithfully transferred without losing
its coherence \cite{Zhao2016SR,Zhao2018SR}. Therefore, a large amount
of research has also been focused on the coherent transfer of cat
state in various physical systems. For example, in Ref. \cite{Jeong2006PRL},
the authors investigate the cat state transfer from a microscopic
system to a macroscopic system. In Ref. \cite{Zeng2020OE}, the authors
investigate the cat state swapping in optomechanical system. In Ref.
\cite{Cosacchi2021PRR}, the authors study the cat state transfer
between a cavity and a solid state quantum-dot system. However, in
all of these physical systems, the research often focus on the cat state transfer caused by the couplings inside quantum systems. The cat state transfer purely caused by the non-Markovian environments has not received enough attention.

\begin{figure}
\begin{centering}
\includegraphics[width=0.7\columnwidth]{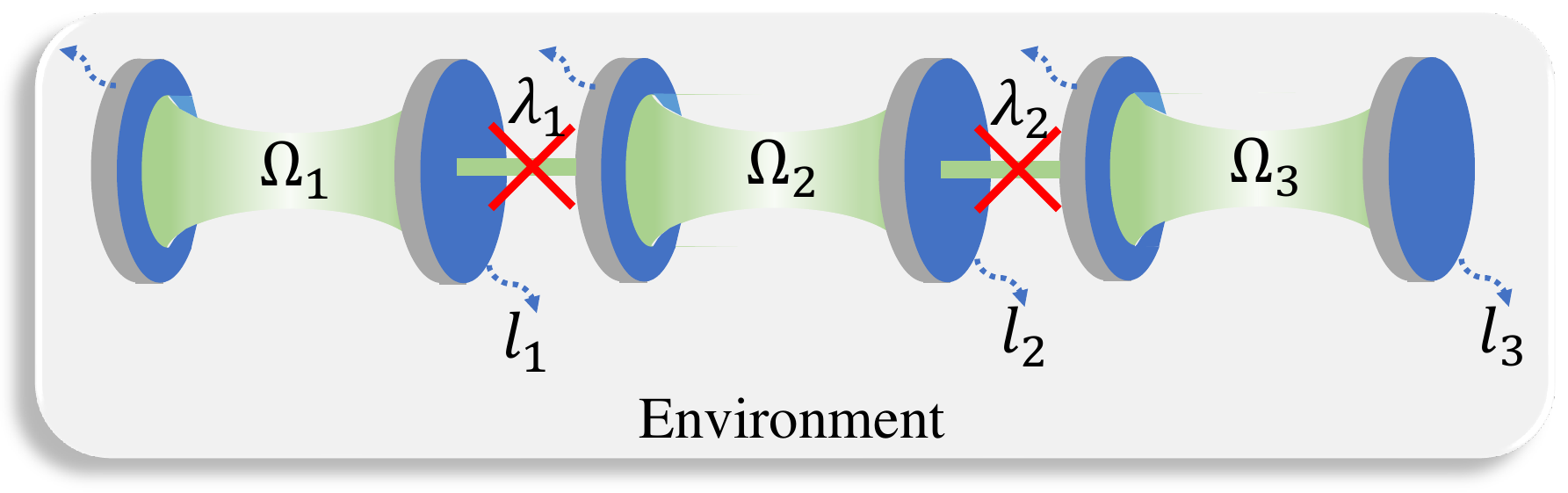}
\par\end{centering}
\caption{\label{Fig:1} Schematic diagram of a coupled $N$-cavity array
(a quantum optical analog of the classical Newton's cradle \cite{Feng2019PRAppl})
interacting with a common environment. In this manuscript, we mainly
focus on the \textquotedblleft no direct coupling\textquotedblright{}
case $\lambda_{i}=0$.}
\end{figure}

In this manuscript, we study the cat state transfer in an optical
analog of Newton's cradle \cite{Feng2019PRAppl} embedded in a non-Markovian
common environment as shown in Fig.~\ref{Fig:1}. Beyond the existing
literature \cite{Cosacchi2021PRR,Dodonov2005JoOBaSO,He2020QIP,Jeong2006PRL,Lidal2020PRA,Prudencio2013IJoQI,Ran2016SR,Teh2018PRA,Zeng2020OE,Liu2020FoP},
our focus in not the transfer induced by the direct couplings between
cavities \cite{Feng2019PRAppl}, but the transfer purely induced by
the non-Markovian environment. Analog to the classical Newton's cradle
composed of several identical balls \cite{Gauld2020PE,Cross2012TPT,Gauld2006SE},
our research is separating the balls and investigate the energy transfer
between the balls through the air resistance. The analytical results
prove that the cat state can indeed transfer purely through environment, and this is a unique phenomenon in non-Markovian regime showing the fundamental difference between Markovian and non-Markovian environments. Besides, the properties of the environment
\cite{Breuer2009PRL,Liu2011NP,Bellomo2007PRL,Liu2007PRA,LoFranco2013IJoMPB,Breuer2016RMP}
is proved to be crucial for the transfer.

To be specific, we first derive the dynamical equations governing
the evolution of the cavity array system in non-Markovian case. Then,
based on these equations, we numerically study the influence of the
memory time of the environment. By monitoring the fidelity and the
Wigner function of the transferred cat state, we point out that the
cat state will be transferred faithfully only in non-Markovian case
when there are no directly couplings. As
a comparison, the transferred state will lose all the coherence in
Markovian case. The mechanism of environment induced transfer is investigated
analytically, and the numerical results also confirm this phenomenon only
exists in non-Markovian case. 

The environment induced cat state transfer reveals that the difference
between non-Markovian and Markovian environment is qualitative (finite
or zero residue coherence) other than quantitative (strong or weak
residue coherence). A simple and straightforward example is that the
weak but non-zero residue coherence in non-Markovian case keeps the
possibility for future distillation \cite{Yu2004PRL,Yu2009S}. Last but not least, we investigate
how to transfer the cat state into a desired cavity by manipulating
the coupling strengths between cavities. By tuning the coupling strengths,
one can transfer the cat state into a particular cavity.

\section{\label{sec:2}Optical Newton's cradle in non-Markovian environment}

\subsection{Quantum mechanical description of optical Newton's cradle}

We consider a coupled cavity array (a quantum version of Newton's
cradle \cite{Feng2019PRAppl}) interacting with a non-Markovian common
environment as shown in Fig.~\ref{Fig:1}. The Hamiltonian of
such a system can be described by
\begin{equation}
H=H_{{\rm S}}+H_{{\rm B}}+H_{\mathrm{int}},\label{Htot}
\end{equation}
with the Hamiltonian of the cavity array system \cite{Liao2010PRA,Meher2017SR,Dubyna2020QST,Saxena2023NC}
\begin{equation}
H_{{\rm S}}=\sum\limits _{i=1}^{N}\Omega_{i}a_{i}^{\dagger}a_{i}+\sum\limits _{i=1}^{N}\lambda_{i}(a_{i}^{\dagger}a_{i+1}+a_{i}a_{i+1}^{\dagger}),\label{H_sys}
\end{equation}
where $\Omega_{i}$ $(i=1,2,...N)$ are the eigen-frequencies for
each cavity in the array, and $\lambda_{i}$ are the coupling strengths
between $i^{{\rm th}}$ cavity and $(i+1)^{{\rm th}}$ cavity. An
open boundary condition is chosen here, namely the first and the last
cavity form two open ends ($\lambda_{N}=0$).

The Hamiltonian in Eq.~(\ref{H_sys}) actually represents a huge
category of quantum system not limited to ``cavities''. For example,
wave-guide system \cite{Feng2019PRAppl}, LC oscillator in superconducting
circuits \cite{You2011N} and many other bosonic systems modeled by
quantum harmonic oscillators \cite{An2007PRA} can be also interpreted
as $H_{{\rm S}}$. Here, we use cavity array as an example to present
our study, but the conclusion should apply to all similar systems.

We assume that all the cavities are coupled to a bosonic common environment
\begin{equation}
H_{{\rm B}}=\sum\limits _{j}\omega_{j}b_{j}^{\dagger}b_{j},\label{H_bath}
\end{equation}
where $b_{j}$ are the annihilation operator of the $j^{{\rm th}}$
mode. The interaction Hamiltonian between the system and the environment under rotating wave approximation 
can be formally written as
\begin{equation}
H_{\mathrm{int}}=\sum\limits _{j}(g_{j}b_{j}^{\dagger}A+g_{j}^{\ast}b_{j}A^{\dagger}),
\end{equation}
where the operator $A=\sum\nolimits _{i=1}^{N}l_{i}a_{i}$ represents
a collective coupling operator between the environment and all the
cavities. For simplicity, $l_{i}$ are always set to be 1, except
in the discussion in Sec.~\ref{sec:Mcav}.

\subsection{Derivation of dynamical equations}

By using the coherent state $|z\rangle=\otimes\prod_{j}|z_{j}\rangle$
representation of the environment, one can define a stochastic state
vector $|\psi(t,z)\rangle \equiv \langle z|\psi_{tot}(t)\rangle$ living in
the Hilbert space of the system only. Here, $|\psi_{tot}(t)\rangle$
is the total state vector describing the state of system plus environment,
and $|z\rangle$ is a collective coherent state of all the environmental
modes, whose $j^{{\rm th}}$ mode is in the coherent state $|z_{j}\rangle$.
Based on the fact that the total state vector $|\psi_{tot}(t)\rangle$
should satisfy the Schr\"{o}dinger equation $\frac{d}{dt}|\psi_{tot}(t)\rangle=-iH_{I}|\psi_{tot}(t)\rangle$
($H_{I}$ is the Hamiltonian in the interaction picture and we set
$\hbar=1$ here and throughout the paper), the stochastic state vector
should satisfy the following dynamical equation called non-Markovian
quantum state diffusion (NMQSD) equation \cite{Diosi1997PLA,Diosi1998PRA,Strunz1999PRL,Yu1999PRA,Zhao2011PRA,Zhao2012PRA,Zhao2017AP,Zhao2019OE,Luo2023arXiv,Shi2024PRA,Luo2024PRA},
\begin{equation}
\frac{\partial}{\partial t}|\psi(t,z)\rangle=\left[-iH_{{\rm S}}+Az_{t}^{\ast}-A^{\dagger}\int\nolimits _{0}^{t}dsK(t,s)\frac{\delta}{\delta z_{s}^{\ast}}\right]|\psi(t,z)\rangle,\label{eq:QSD}
\end{equation}
where $K(t,s)=\sum_{j}|g_{j}|^{2}e^{-i\omega_{j}(t-s)}$ is the correlation function, and $z_{t}^{\ast}=-i\sum_{j}g_{j}z_{j}^{\ast}e^{i\omega_{j}t}$
is a complex stochastic noise satisfying $M[z_{t}]=M[z_{t}z_{s}]=0$,
$M[z_{t}^{\ast}z_{s}]=K(t,s)$. Here $M[\cdot]\equiv\int\frac{dz^{2}}{\pi}e^{-|z|^{2}}[\cdot]$
denotes the statistical average over the noise $z_{t}^{\ast}$. If
we define the environment operator as $B(t)=\sum_{j}g_{j}^{*}b_{j}e^{-i\omega_{j}t}$,
the correlation function can be also interpreted as 
\begin{equation}
K(t,s)=\langle B(t)B^{\dagger}(s)\rangle=\sum_{j}|g_{j}|^{2}e^{-i\omega_{j}(t-s)},
\end{equation}
where the summation over $j$ can be replaced by an integral over
the frequency $\omega$ as $K(t,s)=\int_{0}^{\infty}d\omega g(\omega)e^{-i\omega(t-s)}$.
The distribution function $g(\omega)$ is called the spectrum density
of the environment. 

In order to solve Eq.~(\ref{eq:QSD}), we replace the functional derivative
$\frac{\delta}{\delta z_{s}^{\ast}}$ by an operator $O(t,s,z^{*})$ as,
\begin{equation}
\frac{\delta}{\delta z_{s}^{\ast}}|\psi(t,z)\rangle=O(t,s,z^{\ast})|\psi(t,z)\rangle.
\end{equation}
Then, Eq.~(\ref{eq:QSD}) can be rewritten as 
\begin{equation}
\frac{\partial}{\partial t}|\psi(t,z)\rangle=\left[-iH_{S}+Az_{t}^{*}-A^{\dagger}\bar{O}(t,z^{*})\right]|\psi(t,z)\rangle,\label{eq:QSD2}
\end{equation}
where $\bar{O}(t,z^{*})\equiv\int_{0}^{t}dsK(t,s)O(t,s,z^{*})$ and
the initial condition $O(t,s=t,z^{*})=A$ should be satisfied. Here,
the operator $O$ is replaced by a formally time-local operator $\bar{O}$,
which is a time integration over all the time points $s$ ($s\leq t)$.
The impact of the evolution history (non-Markovian effect) is reflected
in this time-domain convolution.

Now, the key to solving the dynamic equation (\ref{eq:QSD2}) is the
operator $O$. According to the consistency condition $\frac{\delta}{\delta z_{s}^{\ast}}\frac{\partial}{\partial t}|\psi(t,z)\rangle=\frac{\partial}{\partial t}\frac{\delta}{\delta z_{s}^{\ast}}|\psi(t,z)\rangle$
\cite{Yu1999PRA,Diosi1998PRA}, $O$ should satisfy the equation 
\begin{equation}
\frac{\partial}{\partial t}O=[-iH_{{\rm S}}+Az_{t}^{\ast}-A^{\dagger}\bar{O},O]-A^{\dagger}\frac{\delta}{\delta z_{s}^{\ast}}\bar{O}.\label{EQ_O}
\end{equation}
From Eq.~(\ref{EQ_O}), the exact $O$ operator of this model can
be derived as
\begin{equation}
O(t,s)=\sum\limits _{i=1}^{N}f_{i}(t,s)a_{i},\label{Ansatz_O}
\end{equation}
where $f_{i}(t,s)$ are time-dependent coefficients satisfying the
following equations,
\begin{equation}
\frac{\partial}{\partial t}f_{i}(t,s) =i\Omega_{i}f_{i}(t,s)+i[\lambda_{i}f_{i+1}(t,s)+\lambda_{i-1}f_{i-1}(t,s)]
+\sum\limits _{j=1}^{N}f_{j}(t,s)F_{i}(t),\label{df}
\end{equation}
and $F_{i}(t)=\int_{0}^{t}dsK(t,s)f_{i}(t,s)$ $(i=1,2,..,N)$, with
the initial conditions 
\begin{equation}
f_{i}(t=s,s)=1.\label{eq:ini}
\end{equation}
With the exact solution of operator $O$ in Eq.~(\ref{Ansatz_O}),
one can numerically simulate Eq.~(\ref{eq:QSD2}) to obtain the reduced
density operator of the system by taking the statistical average over
the stochastic wave function $|\psi(t,z)\rangle$ as
\begin{equation}
\rho(t)=M[|\psi(t,z)\rangle\langle\psi(t,z)|]=\int\frac{dz^{2}}{\pi}e^{-|z|^{2}}|\psi(t,z)\rangle\langle\psi(t,z)|.\label{eq:rho}
\end{equation}
Alternatively, one can also derive the corresponding master equation
from Eq.~(\ref{eq:rho}) by taking time-derivative on both sides.
Then, using the Novikov theorem \cite{Yu1999PRA,Zhao2012PRA}, one
can obtain the following master equation
\begin{equation}
\frac{d}{dt}\rho=-i\left[H_{{\rm S}},\rho\right]+\left\{ \sum_{k}F_{k}(t)\left[A,\rho a_{k}^{\dagger}\right]+h.c.\right\} .\label{eq:MEQ}
\end{equation}

We would like to emphasize that the NMQSD equation (\ref{eq:QSD}),
as well as the master equation (\ref{eq:QSD2}), are derived directly
from the microscopic Hamiltonian (\ref{Htot}) without any approximation. The detailed derivation is presented in Appendix~\ref{sec:App1}.
It is the exact dynamical equation governing the dynamics of the cavity
array coupled to the environment, particularly in the non-Markovian
regime. This method has been widely studied and used in various physical
systems \cite{Chen2022JCP,Flannigan2022PRL,Qi2021PRA,Turkeshi2021PRB,Ren2020PRA,Wang2020PRA,Luoma2020PRL,Ma2020PRA,Gao2019JCP,Link2019PRA,Zhao2022PRA}.

In the derivation above, we assume the temperature is zero, i.e.,
the initial state of the environment is the vacuum state. However,
the finite temperature case only slightly increase the complexity
of the equations without changing the physical performance we focused
on (see discussion in Sec.~\ref{sec:CatTrans}). In Appendix~\ref{sec:AppA},
the NMQSD equation and the master equation for finite temperature
case are derived.

\subsection{Spectrum density and correlation function of the environment}

The environmental impact on the dynamics of the optical Newton's cradle
is reflected on the terms $Az_{t}^{\ast}$ and $-A^{\dagger}\bar{O}$
in Eq. (\ref{eq:QSD2}). If these two terms are zero, the equation
is reduced to $\partial_{t}|\psi(t,z)\rangle=-iH_{{\rm S}}|\psi(t,z)\rangle$,
which is the Schr\"{o}dinger equation for the closed system. 
The non-Markovian properties are reflected by the correlation function
$K(t,s)$ in Eq.~(\ref{eq:QSD}) through the time-domain convolution
in $\bar{O}$. An evolution without memory effect will be represented
by a $\delta$-type correlation function as $K(t,s)=\Gamma\delta(t,s)$,
which means the evolution at time point ``$t$'' is independent
of the behavior at any previous time point ``$s$''. As a result,
the operator $\bar{O}$ is reduced to $\bar{O}=\frac{\Gamma}{2}A$
due to the initial condition $O(t=s,s,z^{*})=A$. Then, Eq.~(\ref{eq:QSD2})
is reduced to the commonly used Markovian quantum trajectory equation
\cite{Dalibard1992PRL,Gisin1993JoPAaG}, and Eq.~(\ref{eq:MEQ})
is reduced to the standard Lindblad master equation \cite{Yu1999PRA}
with constant coefficients.

The results presented here are independent of a specific form of the
correlation functions $K(t,s)$. Therefore, Eq.~(\ref{eq:QSD2})
and Eq.~(\ref{eq:MEQ}) are applicable to arbitrary correlation functions.
One can definitely use any correlation function either predicted from
microscopic models or directly measured from experiments. However, in
the numerical simulation in Sec.~\ref{sec:CatTrans}, we choose the Lorentzian spectrum density
\begin{equation}
g(\omega)=\frac{\Gamma\gamma^{2}/2\pi}{(\omega-\Delta)^{2}+\gamma^{2}},\label{eq:Lorentz}
\end{equation}
corresponding to the Ornstein-Uhlenbeck (O-U) type correlation function
\begin{equation}
K(t,s)=\frac{\Gamma\gamma}{2}e^{-(\gamma+i\Delta)|t-s|}.\label{eq:CF}
\end{equation}
In Eq.~(\ref{eq:Lorentz}) and (\ref{eq:CF}), the parameter $\tau=1/\gamma$
indicates the memory time of the environment, $\Delta$ is the central
frequency of the environment, and $\Gamma$ is a global dissipation
rate. The Lorentzian spectrum density of the environment has been
widely used in the research on optical systems \cite{Zhao2011PRA,Mu2016PRA,Zhao2019OE}
and other systems \cite{Tu2008PRB,Zhao2012PRA,Shi2013PRA}. The reason
we choose O-U correlation function is because it is easy to observe
the transition from non-Markovian regime to Markovian regime. If the
memory time $\tau=1/\gamma$ is very small in Eq.~(\ref{eq:CF}),
$K(t,s)$ is approximately reduced to $K(t,s)\approx\Gamma\delta(t,s)$,
which means the environment is reduced to a Markovian environment.

It is worth to note the NMQSD equation and master equation derived in this paper are applicable to arbitrary noises. The procedure of numerically generating arbitrary noises can be found in Ref.~\cite{Zhao2022PRA}.

\subsection{Numerical methods: Pure state trajectories and master equation}

To numerically study the dynamics of the optical Newton's cradle,
one can either simulate Eq.~(\ref{eq:QSD2}) and Eq.~(\ref{eq:MEQ}).
Both of them will produce identical results \cite{Yu1999PRA}. However,
for a multipartite continuous variable system like the cavity array,
simulating the density operator $\rho$ is more difficult than
simulating the pure state vector $|\psi(t,z)\rangle$. Since the cat
state is not a Gaussian state, we should use a cut-off $N_c$ on the Fock
basis to a certain photon number $N_c$  for each cavity. The memory usage for storing a state vector containing
three cavities is $\sim N_{c}^{3}$, but the memory usage of storing
a density operator is $\sim N_{c}^{6}$. Besides, the non-Markovian
effect make all the time points ``$s$'' in the history may have
a potential contribution. If the time evolution is divided into 1000
steps, the total memory consumption for density operator approach
is at least $1000N_{c}^{6}$. For $N_{c}=10$ which is far below the
precision requirements for simulating coherent states, the memory
usage for storing a single density operator is 1 Gigabytes. For $N_{c}=20$
which can provide a moderate precision, the usage dramatically increases
to 64 Gigabytes, which is far beyond a typical desktop computer. As
a comparison, the memory usage for state vector is only about $1000N_{c}^{3}$
(about 8 Megabytes). Although the discussion above is merely a rough estimation of the memory usage, one can conclude that the NMQSD equation
(\ref{eq:QSD2}) has a great computational advantage.

Although the master equation lose the competition in numerical simulation,
it has its own advantages. Unlike the stochastic state vector that
has an unclear and controversial physical interpretation \cite{Diosi2008PRL,Wiseman2008PRL},
the physical meaning of density operator is clear. Therefore, the
master equation is helpful for analytical investigation of the physical process.
In Sec.~\ref{sec:Mechanism}, the mechanism of environment induced
cat state transfer is just revealed by analyzing the master equation.
In summary, Eq.~(\ref{eq:QSD2}) and Eq.~(\ref{eq:MEQ}) have their
own advantages, so we have derived both equations and utilize their
strengths in different scenarios.

\section{\label{sec:CatTrans}Environment induced cat state transfer}

In this section, we will focus on the cat state transfer based on
the non-Markovian dynamical equations derived in Sec.~\ref{sec:2}.
We assume the initial state of the $1^{{\rm st}}$ cavity is prepared
in a superposition of two coherent states $|\alpha\rangle$ and $|-\alpha\rangle$,
i.e., the cat state 
\begin{equation}
|\psi_{\mathrm{cat}}\rangle=1/\sqrt{\mathcal{N}}(|\alpha\rangle+|-\alpha\rangle),
\end{equation}
where $\mathcal{N}$ is a normalization factor. The other two cavities are in
the vacuum state, i.e., $|\psi(0)\rangle=|\psi_{\mathrm{cat}}\rangle_{1}\otimes|0\rangle_{2}\otimes|0\rangle_{3}\otimes\cdots\otimes|0\rangle_{N}$.
The fidelity
$\mathcal{F}_{i}$ is used to quantify the performance of the cat state transfer.
Since the cat state may rotate in time evolution, so all the rotated cat state $|\psi_{\mathrm{cat}}^{(R)}(\theta)\rangle=1/\sqrt{\mathcal{N}}(|\alpha e^{-i\theta}\rangle+|-\alpha e^{-i\theta}\rangle)$
with a dynamical phase angle $\theta$ can be considered as a successful
transfer of the cat state. Therefore, we define the fidelity of the
transferred state as
\begin{equation}
\mathcal{F}_{i}=\max_{\{\theta\}}[\langle\psi_{\mathrm{cat}}^{(R)}(\theta)|\rho_{i}(t)|\psi_{\mathrm{cat}}^{(R)}(\theta)\rangle]\;(i=2\;{\rm to}\;N),
\end{equation}
where $\rho_{i}(t)$ are the reduced density matrices for the $i^{{\rm th}}$
cavity, and the fidelity is defined as the maximum fidelity between
$\rho_{i}(t)$ and $|\psi_{\mathrm{cat}}^{(R)}(\theta)\rangle$ over all possible
phase factor $\theta$.

\subsection{\label{subsec:EAT}Memory effect}

\begin{figure}
\begin{centering}
\includegraphics[width=0.7\columnwidth]{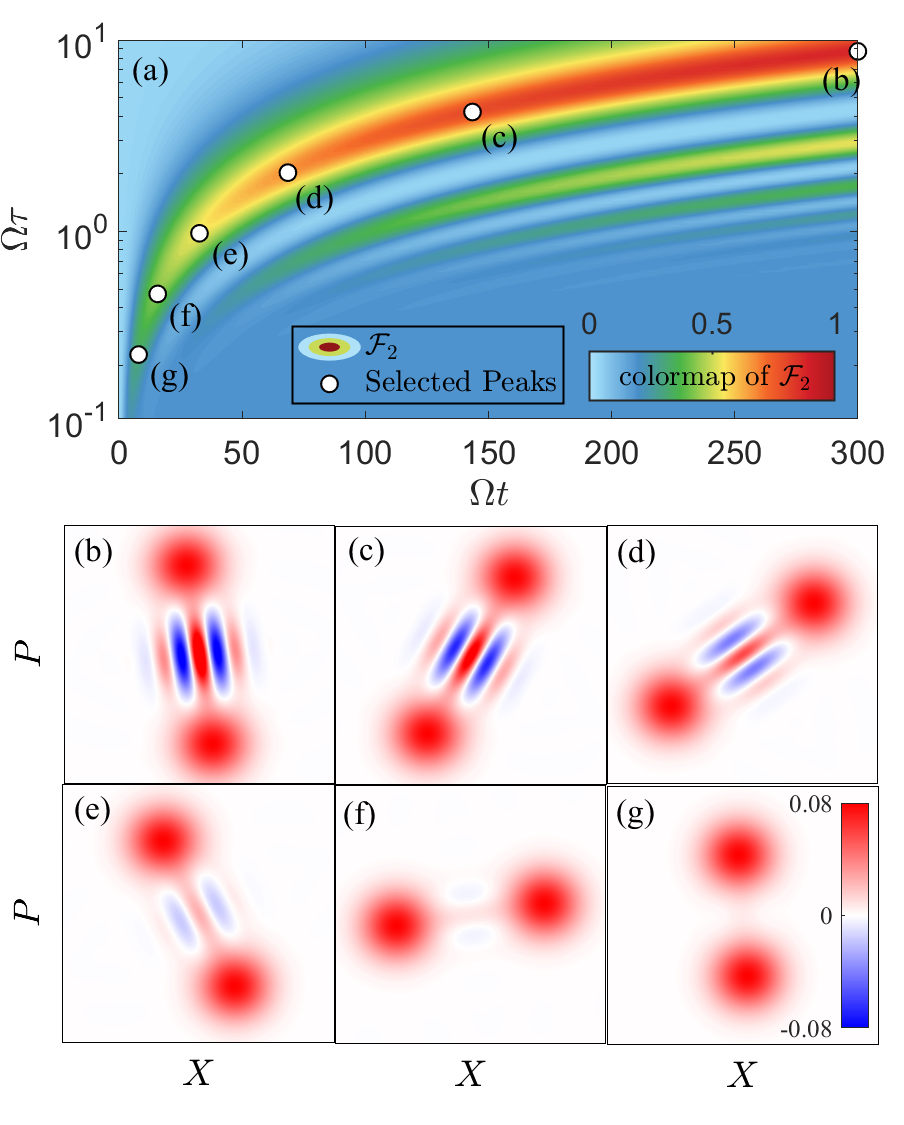}
\par\end{centering}
\caption{\label{Fig:2 2cav}Cat state transfer in two cavities ($N=2$) without
direct couplings $(\lambda_{i}=0)$. (a) Time evolution of the transferred
fidelity $\mathcal{F}_{2}$. The transition from Markovian to non-Markovian
regime is reflected by the increasing $\tau$ in the $y$-axis. (b)-(g)
are the Wigner functions at the positions pointed by the markers in
(a), respectively. The initial size of the cat state is $\alpha=2$.
The other parameters are $\Omega_{1}=\Omega_{2}=\Omega=1$, $\Delta=10$,
$\Gamma=1$. }
\end{figure}

It is not surprising that the cat state can be transferred when the
couplings between neighboring cavities are finite ($\lambda_{i}>0$)
\cite{Feng2019PRAppl}. In this manuscript, the most interesting result is that a high fidelity cat state
transfer can be still achieved without direct couplings ($\lambda_{i}=0$
for all $i$). We first consider a simple case with only two coupled cavities ($N=2$). In Fig.~\ref{Fig:2 2cav},
the time evolution of the fidelity $\mathcal{F}_{2}$ is plotted as
a function of the memory time $\tau=1/\gamma$. When the memory time
is sufficiently long ($\tau$ is large), it is clear that the cat state
has been successfully transferred into the other cavity with a high
fidelity $\mathcal{F}_{2}\approx1$. Since the direct coupling is absent, the common environment
is the only connection among cavities. Therefore, such a cat state
transfer is purely induced by the non-Markovian environment. In this
sense, the non-Markovian properties of the environment must be essential
to the fidelity of the transfer. From Fig.~\ref{Fig:2 2cav}~(a),
the fidelity of the transfer $\mathcal{F}_{2}$ decreases when the
memory time $\tau$ decreases, which implies the high fidelity transfer
only occurs in non-Markovian regime.

A deeper investigation shows that the cat state transfer in Markovian
and non-Markovian regimes can be fundamentally different. To show
this difference, we select several peaks (the maximum $\mathcal{F}_{2}$
can be achieved for given $\tau$) in Fig.~\ref{Fig:2 2cav}~(a)
and plot the Wigner functions of the transferred state in subplots
(b) to (g). The Wigner function is another indicator to quantify the
performance of the transfer, because the residue quantum coherence
(superposition) can be reflected by the negative region in the Wigner
functions. In Fig.~\ref{Fig:2 2cav}~(b), the interference fringe
is clear indicating the coherence is maintained perfectly. This is
also confirmed by the high fidelity $\mathcal{F}_{2}\approx1$. As
a comparison, in Fig.~\ref{Fig:2 2cav}~(g), the negative region
disappears which means a complete loss of quantum coherence after
the transfer. Although the fidelity is still finite $\mathcal{F}_{2}>0$,
the quantum information encoded in the cat state is completely lost
and only classical information is left. The subplots (b) to (g) just
show the transition that coherence is gradually lost when the environment
is changing from non-Markovain to Markovian.

It is worth to note that the physics revealed in the Wigner functions
in Fig.~\ref{Fig:2 2cav}~(b) to (g) is far beyond the information
reflected by the fidelity in Fig.~\ref{Fig:2 2cav}~(a). The Wigner
functions show that the influence of the environment can be qualitative
(zero or finite residue coherence) other than quantitative (low or
high fidelity). Similar topics are widely discussed in the research
on entanglement sudden death versus asymptotic decay \cite{Eberly2007S,Yu2004PRL}.
Despite the comprehensive discussion on the fundamental concepts in
quantum mechanics \cite{Yu2009S}, an apparent difference between
zero residue coherence and weak but finite residue coherence is that
the latter case still keeps the possibility for distillation schemes.

\subsection{Central frequency of the spectrum density}

\begin{figure}
	\begin{centering}
		\includegraphics[width=0.7\columnwidth]{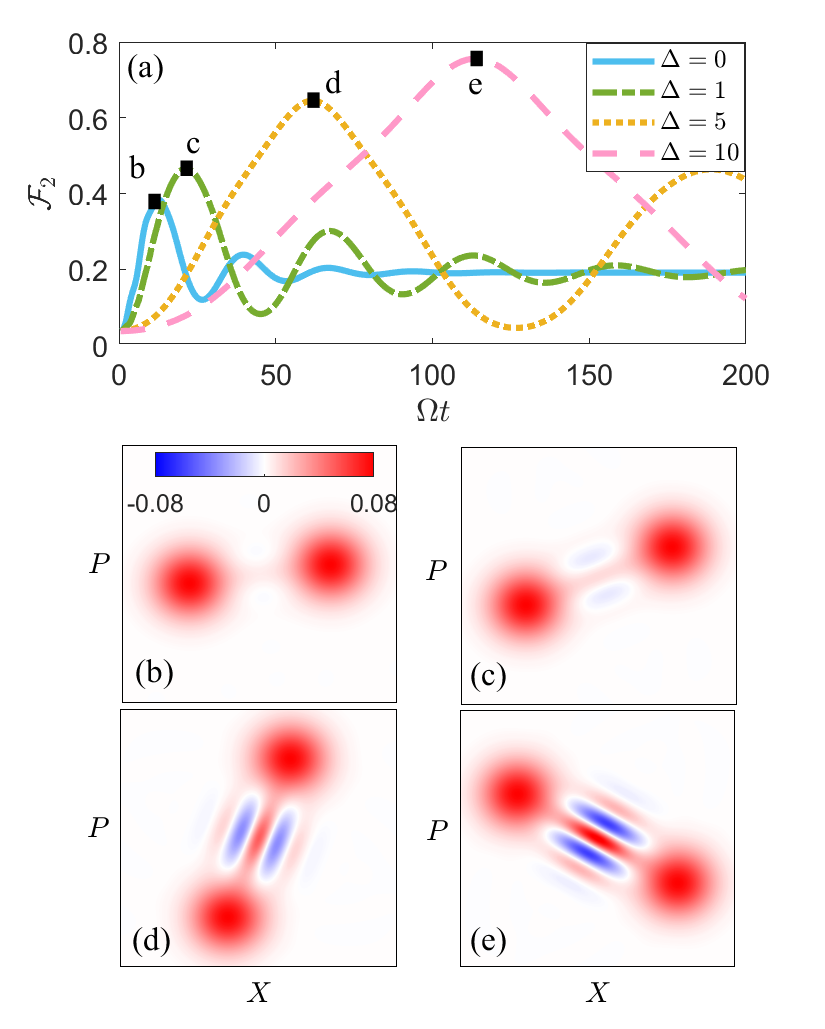}
		\par\end{centering}
	\caption{\label{Fig:3 F2Delta}Influence of the central frequency $\Delta$
		of the environmental density spectrum. (a) Time evolution of the transferred
		fidelity $\mathcal{F}_{2}$. (b) to (e) are the Wigner functions at
		the positions pointed by the markers in (a). The parameters are $\alpha=2$,
		$\gamma=0.3$, $\Omega_{1}=\Omega_{2}=\Omega=1$, $\lambda_{i}=0$,
		$\Gamma=1$. }
\end{figure}

Besides the memory time $\tau$
discussed in Sec.~\ref{subsec:EAT}, the central frequency $\Delta$
is another important environmental parameter which is crucial for the environment induced cat state transfer. In Fig.~\ref{Fig:3 F2Delta}, a higher
central frequency $\Delta$ leads to a larger maximum fidelity $\mathcal{F}_2$, but the
evolution time to achieve maximum $\mathcal{F}_2$ also becomes longer. Similar
to Fig.~\ref{Fig:2 2cav}, the phenomenon that the residue coherence
is completely lost also appears in Fig.~\ref{Fig:3 F2Delta}. When
$\Delta$ is smaller than a certain threshold, the negative region
in the Wigner functions disappears as shown in Fig.~\ref{Fig:3 F2Delta}~(b). This reveals that long memory time $\tau$ is not a sufficient
condition to avoid complete loss of coherence. Even in the case $\tau$
is large enough (as shown in Fig.~\ref{Fig:3 F2Delta}, $\gamma=0.3$
or $\tau=1/\gamma\approx3.33$), coherence can be completely lost
in small $\Delta$ case. Both proper $\tau$ and $\Delta$ are only
necessary conditions to achieve a successful transfer.

In this section, we focus on the case without direct couplings ($\lambda_{i}=0$).
The properties of the environment such as $\tau$ and $\Delta$ are naturally important, since the transfer
is purely induced by the environment. Actually, in the case with direct
couplings ($\lambda_{i}>0$), these environmental parameters are also
crucial for the cat state transfer. In Sec.~\ref{subsec:AppFiniteCouplings}, we
show the non-Markovian effect can significantly enhance the transfer
and cause the revival of the transferred coherence in certain conditions.

\section{\label{sec:Mechanism}Mechanism of environment induced transfer}

The mechanism of the cat state transfer can be revealed by analyzing
the master equation (\ref{eq:MEQ}). In the case $\lambda_{i}=0$,
the first term in Eq.~(\ref{eq:MEQ}) only contains free rotations
$-i\sum_{i=1}^{N}\Omega_{i}\left[a_{i}^{\dagger}a_{i},\rho\right]$
but no interactions between cavities. The indirect interactions between
cavities only originate from the terms $\sum_{k}F_{k}(t)\left[A,\rho a_{k}^{\dagger}\right]+h.c.$.
In the two-cavity case, the coefficients $F_{1}$ and $F_{2}$ are
identical since both the differential equations in (\ref{df}) and
the boundary conditions in (\ref{eq:ini}) are symmetric for $F_{1}$
and $F_{2}$. Therefore, the coefficients can be simplified as $F_{1}=F_{2}=F$
in the symmetric case, leading to the master equation contains the
following indirect interaction term
\begin{equation}
2\Im(F)\left\{ -i\left[\rho,a_{1}^{\dagger}a_{2}+a_{2}^{\dagger}a_{1}\right]\right\}, \label{eq:term}
\end{equation}
where $\Im(F)$ indicates the imaginary part of the coefficient $F$.
If we define $H_{\rm eff}=a_{1}^{\dagger}a_{2}+a_{2}^{\dagger}a_{1}$,
this term can be further written as $2 i \Im(F)\left[H_{\rm eff},\rho\right]$.
It is clear that the contribution of Eq.~(\ref{eq:term}) in the
master equation is equivalent to a direct interaction term in the
Hamiltonian in Eq.~(\ref{H_sys}). The only difference is the former
one originates from the impact of the environment and later one originates
from the direct interaction in the system Hamiltonian. It is well
known that the Hamiltonian in the form $H_{\rm eff}=a_{1}^{\dagger}a_{2}+a_{2}^{\dagger}a_{1}$
(in mathematical perspective, no matter direct or indirect) will cause
a quantum state transfer. This has been widely proved in either previous
papers \cite{Liu2020FoP,Jeong2006PRL,He2020QIP} or the numerical results in Sec.~\ref{subsec:AppFiniteCouplings}.
Therefore, the $\Im(F)$ term in Eq.~(\ref{eq:term}) will cause
a quantum state transfer as if there is a direct interaction in the
system Hamiltonian.

\begin{figure}
	\begin{centering}
		\includegraphics[width=0.7\columnwidth]{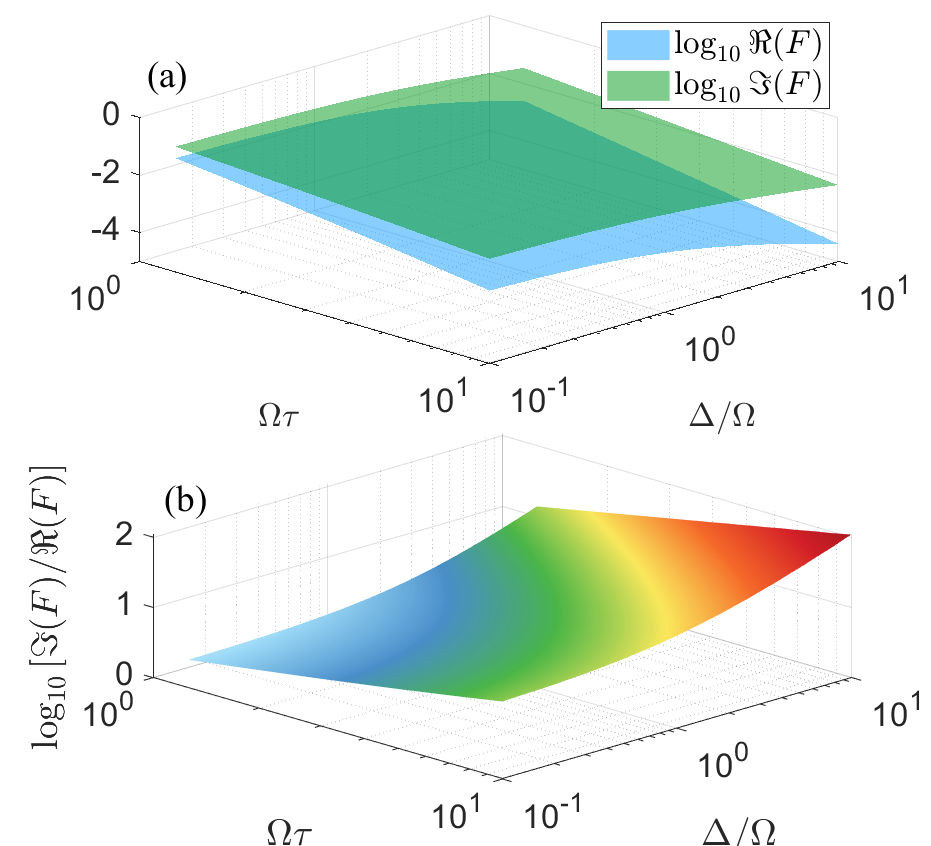}
		\par\end{centering}
	\caption{\label{Fig:4 F}Non-Markovian impact on the long-term ($t\approx\infty$)
		behavior of the coefficient $F$. The real and imaginary part of $F$
		are plotted in (a) with environmental parameters $\Delta$ and $\tau$.
		(b) The ratio of $\Im(F)/\Re(F)$.}
\end{figure}

While the imaginary part $\Im(F)$ leads to an indirect coupling, the
real part $\Re(F)$ corresponds to the traditional dissipation effect
which is similar to the widely studied dissipation term in Lindblad
master equation. The cat state transfer is a combined
effect of both $\Re(F)$ and $\Im(F)$. In Fig.~\ref{Fig:4 F}, the
long-time behavior of $\Re(F)$ and $\Im(F)$ are plotted. The ratio
$\Im(F)/\Re(F)$ somehow reflects which effect is dominant, the dissipation
or the indirect coupling. The numerical results imply that a successful
cat state transfer prefers larger $\tau$ and larger $\Delta$. This
is consistent with the results shown in Fig.~\ref{Fig:2 2cav} and
Fig.~\ref{Fig:3 F2Delta}.

It is worth to note that the $\Im(F)$ term in Eq.~(\ref{eq:term})
only exists in non-Markovian case. One can check that $F_{1}(t)=F_{2}(t)=\Gamma/2$
when $K(t,s)=\Gamma\delta(t,s)$ in a straightforward manner. Namely,
in Markovian case, the $\delta$-type correlation function results
in $\Im(F_{1})=\Im(F_{2})=0$ and the master equation is reduced to
the Lindblad form \cite{Yu1999PRA}. In another word, non-Markovian
correlation function is a necessary condition to produce the indirect
interaction in Eq.~(\ref{eq:term}). From this point of view, we
have mathematically proved that the environment induced cat state
transfer is a purely non-Markovian phenomenon. This also explains
why the cat state can not be transferred in weak Markovian case as
shown in Fig.~\ref{Fig:2 2cav}~(g), because the indirect interaction
$\Im(F)$ is too weak to fight against the dissipation $\Re(F)$.
Quantum coherence is completely lost before it is transferred. Again,
this is qualitatively different from quantitative impact of the environment.

Last but not least, we would like to emphasize that in the finite
temperature case, the conclusion that environment induced transfer
is purely a non-Markovian phenomenon still holds. Similar to the case
in zero temperature case, the coefficients in the $\bar{O}$ operator
become constant real numbers. For more details, see the derivation
in Appendix~\ref{sec:AppA}.

\section{\label{sec:Mcav}Environment induced cat state transfer in multiple
cavities}

\begin{figure}
	\begin{centering}
		\includegraphics[width=0.7\columnwidth]{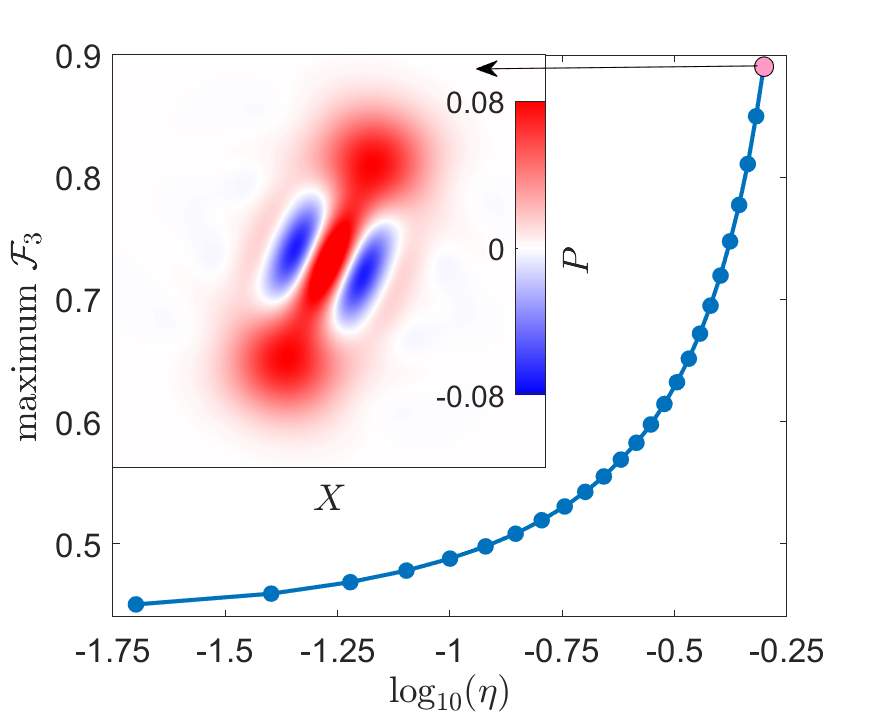}
		\par\end{centering}
	\caption{\label{Fig:5r}Cat state transfer from $1^{{\rm st}}$ cavity to $3^{{\rm rd}}$
		cavity without direct couplings $(\lambda_{i}=0)$. Solid curve indicates
		the maximum fidelity $\mathcal{F}_{3}$ achievable with the asymmetric
		parameter $\eta$. The inset plot is the Wigner function of the $3^{{\rm rd}}$
		cavity when $\log_{10}(\eta)\approx-0.3$ ($\eta=0.5$, $\delta l=0.5$).
		The parameters are $\Omega_{i}=1$, $\gamma=0.1$, $\Delta=5$,}
\end{figure}

In Sec.~\ref{sec:CatTrans}, we only show the two-cavity case ($N=2$)
as an example. In this section, we will focus on the case that multiple
cavities are coupled to the environment. Taking three-cavity case
($N=3$) as an example, the $2^{{\rm nd}}$ and $3^{{\rm rd}}$ cavities
are in the symmetric positions. Due to this symmetric property in
Eq.~(\ref{eq:MEQ}) as well as in its initial conditions, the solution
of $\rho^{(2)}(t)$ will be identical to $\rho^{(3)}(t)$. In order
to transfer the cat state into a desired cavity purely through the assistance
of environment, the coupling strengths to the environment $l_{1}$,
$l_{2}$, and $l_{3}$ must be different. Here, we assume $l_{1}=1$,
$l_{2}=1-\delta l$, and $l_{3}=1+\delta l$, where $\delta l$ describes
the difference of the coupling strengths between the $2^{{\rm nd}}$
and $3^{{\rm rd}}$ cavities. An asymmetric parameter can be defined
as 
\begin{equation}
\eta=\frac{l_{3}-l_{2}}{l_{2}+l_{3}},
\end{equation}
to measure the asymmetry of the two coupling strengths $l_{2}$ and
$l_{3}$.

It is straightforward to draw a conclusion for the two limiting cases
$\eta=-1$ and $\eta=1$. For $\eta=-1$, $l_{3}=0$, the model is
reduced into $N=2$ case discussed in Sec.~\ref{sec:CatTrans}, because
the $3^{{\rm rd}}$ cavity is decoupled from the environment thus
losing all the connections to other cavities (no directly coupling
$\lambda_{i}=0$). Similarly, in the case $\eta=1$, the $2^{{\rm nd}}$
cavity is decoupled. One may expect that a high-fidelity transfer to the $3^{{\rm rd}}$
cavity needs $\eta\approx1$ or $\delta l\approx1$ to almost fully
decouple the $2^{{\rm nd}}$ cavity. However, the numerical results
in Fig.~\ref{Fig:5r} show a mild asymmetry can just ensure a high-fidelity
transfer. The pink marker in the main plot shows that a high fidelity
$\mathcal{F}_{3}\approx0.9$ can be achieved for $\eta=0.5$, i.e.,
$\delta l=0.5$, far from the limiting case $\eta=1$. Besides, the
fidelity increase rapidly with the increase of $\eta$. So, a decent
transfer can be achieved when there is a tiny asymmetry. 

In the finite direct coupling case, one can similarly tune the direct
couplings $\lambda_{i}$ and introduce an asymmetry to transfer the
cat state into a desired cavity. This is discussed in details in Appendix~\ref{subsec:AssCoup}.
By tuning $\lambda_{i}$, similar behaviors in the classical Newton's
cradle (only the balls on the ends will be lifted) can be reproduce
in the quantum version analog.

\section{Cat state transfer with finite direct couplings \label{subsec:AppFiniteCouplings}}

\begin{figure}
	\begin{centering}
		\includegraphics[width=0.7\columnwidth]{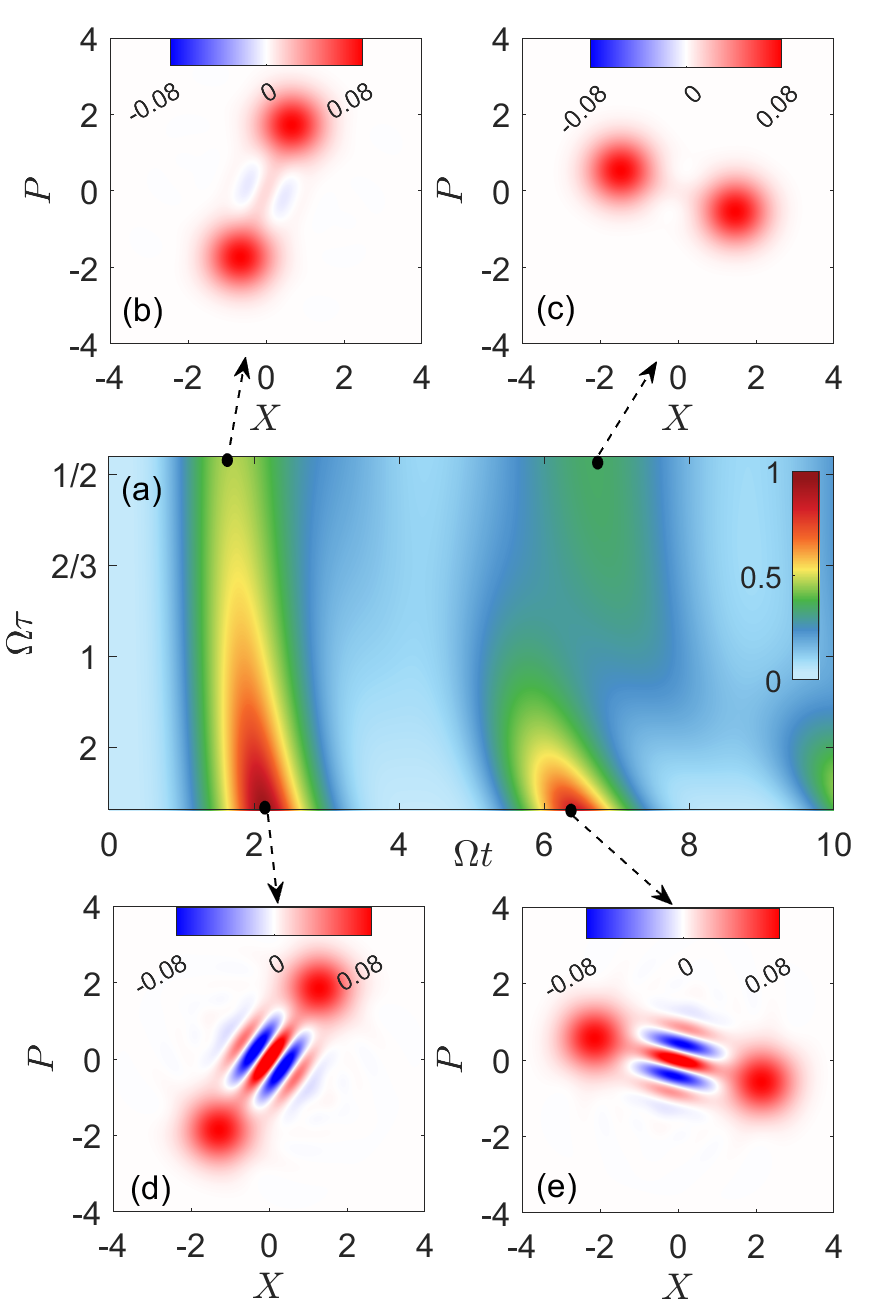}
		\par\end{centering}
	\caption{\label{Fig:6}Memory effect on the time evolution of the transferred
		fidelity $\mathcal{F}_{3}$ with finite $\lambda_{i}$. Subplot (a)
		shows $\mathcal{F}_{3}$ as a function of the memory time $\gamma$
		and the evolution time $\Omega t$. Subplots (b)-(e) show the Wigner
		functions at the positions pointed by the arrows. The parameters are
		$\Omega_{1}=\Omega_{2}=\Omega_{3}=\Omega=1$, $\lambda_{1}=\lambda_{2}=1$,
		$\Delta=0$, and $\Gamma=1$.}
\end{figure}

In the previous sections, we keep our focus on the environment induced cat
state transfer without direct couplings ($\lambda_{i}=0$). However,
the non-Markovian effect also has a significant impact in finite direct
coupling cases ($\lambda_{i}>0$). In this section, we will study
the finite direct coupling case and show the role of non-Markovian
environment in the transfer. Throughout this section, we consider the three-cavity case ($N=3$).

\subsection{\label{subsec:FC_tau}Enhanced transfer by memory effect}

In Fig.~\ref{Fig:6}~(a), we plot the fidelity $\mathcal{F}_{3}$
of the transfer as a function of the memory time $\tau$ and the evolution
time $\Omega t$. It is shown that the cat state can be transferred
into the third cavity with a large fidelity in the non-Markovian case
($\tau$ is large). As a comparison, in the Markovian case ($\tau$
is small), the fidelity of the transfer will be reduced. This is also
directly reflected from the Wigner function of the transferred state
in the subplots. In subplot Fig.~\ref{Fig:6}~(d), the interference
fringe is clear in the Wigner function, while in subplot Fig.~\ref{Fig:6}~(b),
the negative regions are almost invisible. It is known that the quantum
superposition can be reflected by the negative regions in Wigner functions,
so the shrink of the negative region in Fig.~\ref{Fig:6}~(b) indicates
the coherence (superposition) is partially lost. Therefore, the numerical
result shows that only in the non-Markovian regime, strong coherence
can be preserved in the cat state transfer. Another important feature
of the non-Markovian case is the cat state can be transferred back
and forth in the cavity array. In Fig.~\ref{Fig:6}~(a), it is clear
that the fidelity $\mathcal{F}_{3}$ has a revival when $\tau$ is
large. This is the unique feature which can be only observed in non-Markovian
regime. When $\tau<1$ (close to Markovian regime), there is almost
no revival. The Wigner functions plotted in subplots (c) and (e) also
confirm this phenomenon. In Fig.~\ref{Fig:6}~(e), negative region
is clear in the Wigner function, while in Fig.~\ref{Fig:6}~(c),
there is no negative region at all, the coherence (superposition)
is completely lost.

\subsection{\label{subsec:AssCoup}Asymmetric direct couplings}

\begin{figure}
	\begin{centering}
		\includegraphics[width=0.7\columnwidth]{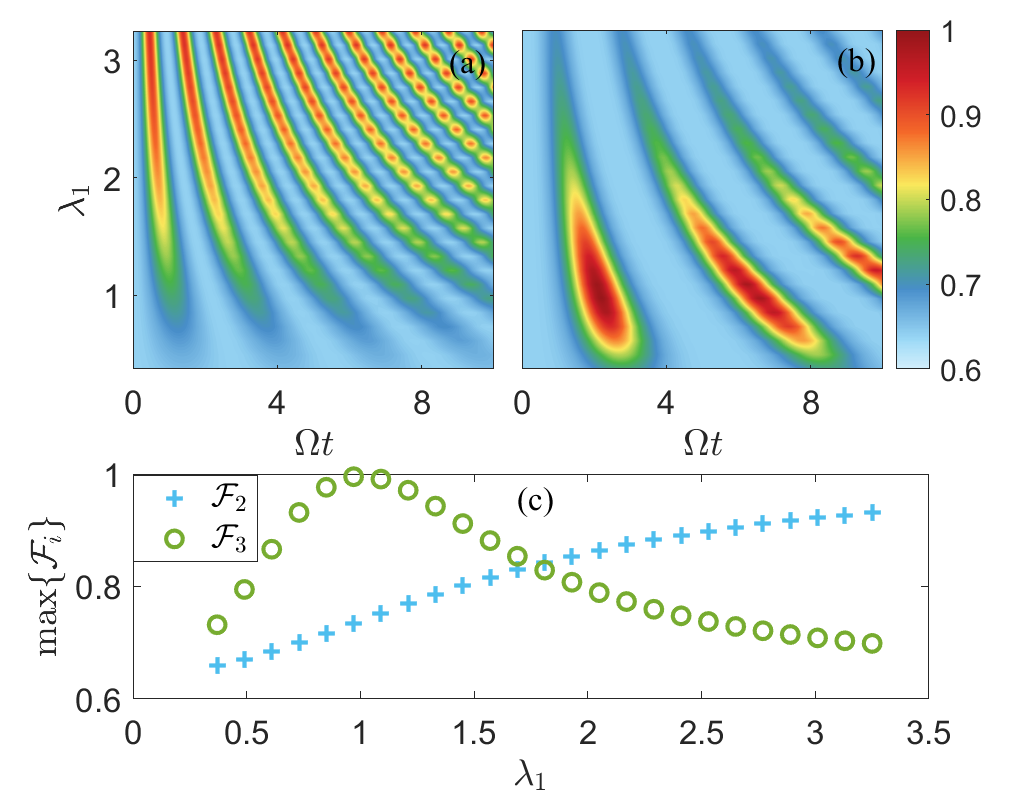}
		\par\end{centering}
	\caption{\label{Fig:7 AsCoup}(a) and (b) are the time evolution of transferred
		fidelities $\mathcal{F}_{2}$ and $\mathcal{F}_{3}$, respectively.
		(c) is the maximum values (over $\Omega t$) can be achieved in the
		time evolution as a function of $\lambda_{1}$. The coupling strengths
		are chosen asymmetrically, while $\lambda_{2}$ is fixed as 1, $\lambda_{1}$
		is varying from 0.25 to 3.25.}
\end{figure}

In the case of finite direct couplings in multiple cavities, the coupling
strengths in a cavity array can be different, either because some
inevitable practical factors or simply because they are intentionally
tuned to be different. Interestingly, in the numerical simulation,
we find the maximum transferred fidelity for the $2^{{\rm nd}}$ cavity
and the $3^{{\rm rd}}$ cavity depends on the choice of the couplings.
In Fig.~\ref{Fig:7 AsCoup}, we plot the time evolution of the transferred
fidelity $\mathcal{F}_{2}$ and $\mathcal{F}_{3}$ with different
coupling strength $\lambda_{1}$ when $\lambda_{2}$ is fixed as 1.
The value of $\lambda_{1}$ can be also regarded as the ratio of $\lambda_{1}/\lambda_{2}$,
indicating a asymmetric parameter. From Fig.~\ref{Fig:7 AsCoup}~(a)
and (b), the cat state tends to be transferred into the $2^{{\rm nd}}$
cavity when $\lambda_{1}$ is large. While $\lambda_{1}$ is small,
the cat state is more likely to be transferred into the $3^{{\rm rd}}$
cavity. This is further illustrated in Fig.~\ref{Fig:7 AsCoup}~(c),
where the maximum transferred fidelity can be achieved are plotted
as the function of $\lambda_{1}$. It is clear that the peaks of $\mathcal{F}_{2}$
and $\mathcal{F}_{3}$ curves appear at different values of $\lambda_{1}$.

To understand the impact of asymmetric couplings shown in Fig.~\ref{Fig:7 AsCoup},
one can consider the classical Newton's cradle. The middle ball will
keep stand because its position. In the quantum version of Newton's
cradle, the wave function is traveling through three cavities. At
each end, the wave will be reflected, forming a standing wave in the
rope. For a standing wave, the amplitude at some positions can be
always much smaller than the other places. The format and properties
of such a ``standing wave'' traveling in the cavity array are certainly
determined by the boundary condition and the configuration of couplings.
This implies one can control the destination of the transfer in a
cavity array by only tuning the coupling strength. By choosing an
appropriate set of $\lambda_{i}$, the cat state can be precisely
transferred into the desired cavity without worrying about transferring
into another unwanted cavity. This mimic the behavior of classical
Newton's cradle in which only the balls on the ends will be lifted.

\section{\label{sec:4}Conclusion}

In this manuscript, we investigate the cat state transfer in a cavity
array embedded in a non-Markovian common environment. Based on the
master equation derived beyond the Markovian approximation, we numerically
study the cat state transfer in the cavity array. Different from traditional
studies \cite{Feng2019PRAppl,Liu2020FoP,Zeng2020OE}, our focus is the cat state transfer purely induced by non-Markovian
environment without any directly couplings ($\lambda_{i}=0$). We
find that the transfer can be fundamentally different in Markovian
and non-Markovian cases, reflecting by the zero or finite residue
coherence. This extends the traditional understanding that non-Markovian
effect only has quantitative influence (enhance or weaken) on the
dynamics. The results presented here break this understanding and
show the qualitative difference for Markovian and non-Markovian dynamics
(can or cannot transfer cat state). This is proved analytically by
pointing out the necessary condition for environment induced cat state
transfer is the imaginary part of coefficient $F$, which only exists
in non-Markovian case. We hope the results presented in the paper
would be useful to the future study on cat state transfer, particularly
to the investigation of the environmental impacts on the cat state
transfer.

\section*{Appendix}
\appendix

\section{\label{sec:App1}Derivation of master equation}

In this section, we derive the master equation from the NMQSD equation. First, we define the stochastic density operator as $P_{t}\equiv|\psi(t,z^{*})\rangle\langle\psi(t,z)|$, the density operator is the statistical mean over all possible stochastic density operators as shown in Eq.~(\ref{eq:rho}). Taking the time-derivative
to $\rho$ and notice the NMQSD Eq.~(\ref{eq:QSD2}), we obtain
\begin{eqnarray}
	\frac{d}{dt}\rho & = & \frac{d}{dt}M\{P_{t}\}\nonumber \\
	& = & M\{[\frac{d}{dt}|\psi(t,z^{*})\rangle\langle\psi(t,z)|]\}+M\{[|\psi(t,z^{*})\rangle\frac{d}{dt}\langle\psi(t,z)|]\}\nonumber \\
	& = & M\{[-iH_{S}+Az_{t}^{*}-A^{\dagger}\bar{O}]P_{t}\}+M\{P_{t}[iH_{S}+A^{\dagger}z_{t}-\bar{O}^{\dagger}A]\}.\label{eq:drhos}
\end{eqnarray}
In order to compute the term  $M\{Lz_{t}^{*}P_{t}\}$, we recall the
Novikov theorem, which has been proved in Ref.~\cite{Yu1999PRA},
\begin{equation}
	M\{P_{t}z_{t}\}=M\{\bar{O}P_{t}\},\label{eq:NovikovBoson1}
\end{equation}
\begin{equation}
	M\{z_{t}^{*}P_{t}\}=M\{P_{t}\bar{O}^{\dagger}\}.\label{eq:NovikovBoson2}
\end{equation}
It is not difficult to prove the two relations above. By definition,
\begin{eqnarray}
	M\{P_{t}z_{t}\} & = & {\displaystyle \int}{\displaystyle \prod\nolimits_{k}}dz_{k}^{\ast}dz_{k}\exp(-\sum_{k}z_{k}^{\ast}z_{k})P_{t}(i\sum_{j}g_{j}^{*}e^{-i\omega_{j}t}z_{j})\nonumber \\
	& = & -i\sum_{j}g_{j}^{*}e^{-i\omega_{j}t}z_{j}{\displaystyle \int}{\displaystyle \prod\nolimits_{k}}dz_{k}^{\ast}dz_{k}P_{t}\frac{\partial}{\partial z_{j}^{\ast}}[\exp(-\sum_{k}z_{k}^{\ast}z_{k})].
\end{eqnarray}
Integrating by parts, and noticing that in the polar coordinates,
$[re^{-|r|^{2}}]_{0}^{\infty}=0$, we obtain
\begin{eqnarray}
	M\{P_{t}z_{t}\} & = & i\sum_{j}g_{j}^{*}e^{-i\omega_{j}t}z_{j}{\displaystyle \int}{\displaystyle \prod\nolimits_{k}}dz_{k}^{\ast}dz_{k}\frac{\partial}{\partial z_{j}^{\ast}}P_{t}\exp(-\sum_{k}z_{k}^{\ast}z_{k})\nonumber \\
	& = & i\sum_{j}g_{j}^{*}e^{-i\omega_{j}t}z_{j}{\displaystyle \int}{\displaystyle \prod\nolimits_{k}}dz_{k}^{\ast}dz_{k}(\int ds\frac{\partial z_{s}^{\ast}}{\partial z_{j}^{\ast}}\frac{\delta}{\delta z_{s}^{\ast}})P_{t}\exp(-\sum_{k}z_{k}^{\ast}z_{k})\nonumber \\
	& = & \int ds\sum_{j}\left\vert g_{j}\right\vert ^{2}e^{-i\omega_{j}(t-s)}{\displaystyle \int}{\displaystyle \prod\nolimits_{k}}dz_{k}^{\ast}dz_{k}\exp(-\sum_{k}z_{k}^{\ast}z_{k})\frac{\delta}{\delta z_{s}^{\ast}}P_{t}\nonumber \\
	& = & {\displaystyle \int}{\displaystyle \prod\nolimits_{k}}dz_{k}^{\ast}dz_{k}\exp(-\sum_{k}z_{k}^{\ast}z_{k})\int dsK(t,s)OP_{t}\nonumber \\
	& = & M\{\bar{O}P_{t}\}.
\end{eqnarray}
Similarly, one can prove
\begin{equation}
	M\{z_{t}^{*}P_{t}\}=M\{P_{t}\bar{O}^{\dagger}\}.
\end{equation}
Applying the Novikov theorem Eq. (\ref{eq:NovikovBoson1}) and (\ref{eq:NovikovBoson2})
to Eq. (\ref{eq:drhos}), we obtain the master equation as
\begin{equation}
	\frac{d}{dt}\rho =-i[H_{S},\rho]+\left\{ [A,M\{P_{t}\bar{O}^{\dagger}\}]+h.c.\right\}.\label{eq:MEQ_QSD}
\end{equation}
In our case, the $O$ operator derived in Eq.~(\ref{Ansatz_O}) is independent
of noise variable $z$, $M\{P_{t}\bar{O}^{\dagger}\} = M\{P_{t}\}\bar{O}^{\dagger}=\rho\bar{O}^{\dagger}$ the master equation will reduce to
\begin{equation}
	\frac{d}{dt}\rho=-i[H_{S},\rho]+[A,\rho\bar{O}^{\dagger}]+[\bar{O}\rho,A^{\dagger}].\label{eq:MEQO0}
\end{equation}
This is the master equation~(\ref{eq:MEQ}) in the main text.

\section{\label{sec:AppA}Master equation in finite temperature case}

In the finite temperature case, the initial state of the environment
is the thermal state $\rho_{{\rm B}}(0)=\exp(-\beta H_{{\rm B}})/Z$,
with $Z={\rm Tr}[\exp(-\beta H_{{\rm B}})]$ the partition function
and $\beta=1/k_{B}T$. According to Bose-Einstein statistics, the
occupation number of the $j^{{\rm th}}$ mode is
\begin{equation}
\langle b_{j}^{\dagger}b_{j}\rangle=\bar{n}_{j}=\frac{1}{e^{-\beta\omega_{j}}-1}.\label{eq:nk}
\end{equation}
In order to derive the master equation governing the dynamics of the
cavity array system described by the Hamiltonian (\ref{Htot}) in
the finite temperature case, we introduce a fictitious environment
\cite{Yu2004PRA}
\[
H_{{\rm B^{\prime}}}=\sum_{j}\omega_{j}c_{j}^{\dagger}c_{j},
\]
where $c_{j}$ is the annihilation operator of the $j^{{\rm th}}$
mode of the fictitious environment. Adding the fictitious environment
without interactions with other parts, the total Hamiltonian becomes
\begin{equation}
H=H_{{\rm S}}+H_{{\rm B}}+H_{\mathrm{int}}-H_{{\rm B^{\prime}}}.
\end{equation}
Since the fictitious environment evolves independently, it has no
impact on the evolution of the system. After a Bogoliubov transformation
\begin{equation}
b_{j}=\sqrt{\bar{n}_{j}+1}d_{j}+\sqrt{\bar{n}_{j}}e^{\dagger},
\end{equation}
\begin{equation}
c_{j}=\sqrt{\bar{n}_{j}+1}e_{j}+\sqrt{\bar{n}_{j}}d^{\dagger},
\end{equation}
the Hamiltonian becomes
\begin{eqnarray}
H^{\prime} & = & H_{{\rm S}}+\sum_{j}\omega_{j}d_{j}^{\dagger}d_{j}+\sum_{j}\sqrt{\bar{n}_{j}+1}(g_{j}Ad_{j}^{\dagger}+g_{j}^{*}A^{\dagger}d_{j})\nonumber \\
 &  & -\sum_{j}\omega_{j}e_{j}^{\dagger}e_{j}+\sum_{j}\sqrt{\bar{n}_{j}}(g_{j}Ae_{j}+g_{j}^{*}A^{\dagger}e_{j}^{\dagger}).\label{eq:HFT-1}
\end{eqnarray}
It is easy to check the vacuum state for the system $|0\rangle=|0\rangle_{d}\otimes|0\rangle_{e}$
(satisfying $d_{j}|0\rangle=0$, and $e_{j}|0\rangle=0$) keeps the
original Bose-Einstein distribution in Eq. (\ref{eq:nk}) as
\begin{equation}
\langle0|_{d}\langle0|_{e}b_{j}^{\dagger}b_{j}|0\rangle_{d}|0\rangle_{e}=\bar{n}_{j}.
\end{equation}
Therefore, the finite temperature case is mapped into a zero-temperature
case with two individual environments. Solving the Hamiltonian $H^{\prime}$
with initial state $|0\rangle=|0\rangle_{d}\otimes|0\rangle_{e}$
is equivalent to solving the original Hamiltonian (\ref{Htot}) with
thermal initial state $\rho_{{\rm B}}(0)=\exp(-\beta H_{{\rm B}})/Z$.

Similar to the derivation of Eq.~(\ref{eq:QSD}), one can define
$|\psi_{z,w}\rangle=\langle z,w|\psi_{{\rm tot}}(t)\rangle$, where
$\langle z|$ and $\langle w|$ are the coherent states of bosonic
modes ``$d$'' and ``$e$'', respectively. Then, the NMQSD equation
for Hamiltonian $H^{\prime}$ is given by
\begin{equation}
\frac{\partial}{\partial t}|\psi_{z,w}\rangle =(-iH_{\mathrm{s}}+Az_{t}^{\ast}+A^{\dagger}w_{t}^{\ast})|\psi_{z,w}\rangle
 -A^{\dagger}\int\nolimits _{0}^{t}dsK_{1}(t,s)\frac{\delta|\psi_{z,w}\rangle}{\delta z_{s}^{\ast}}
 -A\int\nolimits _{0}^{t}dsK_{2}(t,s)\frac{\delta|\psi_{z,w}\rangle}{\delta w_{s}^{\ast}},\label{eq:QSDFT}
\end{equation}
where $z_{t}^{\ast}=-i\sum_{j}p_{j}z_{j}^{\ast}e^{i\omega_{j}t}$,
$w_{t}^{\ast}=-i\sum_{j}q_{j}^{\ast}w_{j}^{\ast}e^{-i\omega_{j}t}$
are two statistically independent noises with $p_{j}=\sqrt{\bar{n}_{j}+1}g_{j}$
and $q_{j}=\sqrt{\bar{n}_{j}}g_{j}$ as the effective coupling constants.
The two Gaussian noises satisfy the following relations, 
\begin{align}
M[z_{t}] & =M[z_{t}z_{s}]=0,\quad M[z_{t}z_{s}^{*}]=K_{1}(t,s),\\
M[w_{t}] & =M[w_{t}w_{s}]=0,\quad M[w_{t}w_{s}^{*}]=K_{2}(t,s).
\end{align}
where $K_{1}(t,s)=\sum_{j}|p_{j}|^{2}e^{-i\omega_{j}(t-s)}$ and $K_{2}(t,s)=\sum_{j}|q_{j}|^{2}e^{i\omega_{j}(t-s)}$
are correlation functions for the two effective environments. Similar
to Eq.~(\ref{EQ_O}), the two functional derivatives can be also
replaced by two operators $O_{1}(t,s,z^{\ast},w^{\ast})|\psi_{z,w}\rangle=\frac{\delta}{\delta z_{s}^{\ast}}|\psi_{z,w}\rangle$
and $O_{2}(t,s,z^{\ast},w^{\ast})\psi_{z,w}=\frac{\delta}{\delta w_{s}^{\ast}}|\psi_{z,w}\rangle$.
According to the consistency conditions $\frac{d}{dt}\frac{\delta}{\delta z_{s}^{*}}|\psi_{z,w}\rangle=\frac{\delta}{\delta z_{s}^{*}}\frac{d}{dt}|\psi_{z,w}\rangle$
and $\frac{d}{dt}\frac{\delta}{\delta w_{s}^{*}}|\psi_{z,w}\rangle=\frac{\delta}{\delta w_{s}^{*}}\frac{d}{dt}|\psi_{z,w}\rangle$,
the operators $O_{1}$ and $O_{2}$ should satisfy the following equations
\begin{align}
\frac{\partial}{\partial t}O_{1} & =[-iH_{\mathrm{s}}+Az_{t}^{\ast}+A^{\dagger}w_{t}^{\ast}-A^{\dagger}\bar{O}_{1}-A\bar{O}_{2},O_{1}]
-A^{\dagger}\frac{\delta}{\delta z_{s}^{\ast}}\bar{O}_{1}-A\frac{\delta}{\delta z_{s}^{\ast}}\bar{O}_{2},\label{FT_Eq_O1}\\
\frac{\partial}{\partial t}O_{2} & =[-iH_{\mathrm{s}}+Az_{t}^{\ast}+A^{\dagger}w_{t}^{\ast}-A^{\dagger}\bar{O}_{1}-A\bar{O}_{2},O_{2}]
-A^{\dagger}\frac{\delta}{\delta w_{s}^{\ast}}\bar{O}_{1}-A\frac{\delta}{\delta w_{s}^{\ast}}\bar{O}_{2}.\label{FT_Eq_O2}
\end{align}
where $\bar{O}_{i}=\int_{0}^{t}K_{i}(t,s)O_{i}(t,s,z^{\ast},w^{\ast})ds$
($i=1,2$). According to Eqs.~(\ref{FT_Eq_O1}-\ref{FT_Eq_O2}),
the exact $O$ operators for the $N$-cavity model can be determined
as 
\begin{align}
O_{1}(t,s,w^{\ast}) & =\sum\limits _{i=1}^{N}x_{i}(t,s)a_{i}+\int_{0}^{t}x^{\prime}(t,s,s^{\prime})w_{s^{\prime}}^{\ast}ds^{\prime},\label{FTO1}\\
O_{2}(t,s,z^{\ast}) & =\sum\limits _{i=1}^{N}y_{i}(t,s)a_{i}^{\dagger}+\int_{0}^{t}y^{\prime}(t,s,s^{\prime})z_{s^{\prime}}^{\ast}ds^{\prime},\label{FTO2}
\end{align}
while the coefficients satisfy the following equations 
\begin{align}
 \frac{\partial}{\partial t}x_{i}(t,s)
 =&  i\Omega_{i}x_{i}(t,s)+i[\lambda_{i}x_{i+1}(t,s)+\lambda_{i-1}x_{i-1}(t,s)]\nonumber \\
&  +\sum\limits _{j=1}^{N}x_{j}(t,s)[Y_{j}(t)+X_{i}(t)]-Y^{\prime}(t,s),\label{eq:ptxi}
\end{align}
\begin{align}
 \frac{\partial}{\partial t}y_{i}(t,s)
= &  -i\Omega_{i}y_{i}(t,s)-i[\lambda_{i}y_{i+1}(t,s)+\lambda_{i-1}y_{i-1}(t,s)]\nonumber \\
 & -\sum\limits _{j=1}^{N}y_{j}(t,s)X_{j}(t)-\sum\limits _{j=1}^{N}y_{j}(t,s)Y_{i}(t)-X^{\prime}(t,s),
\end{align}
\begin{equation}
\frac{\partial}{\partial t}x^{\prime}(t,s,s^{\prime})=\sum\limits _{j=1}^{N}x_{j}(t,s)X^{\prime}(t,s^{\prime}),
\end{equation}
\begin{equation}
\frac{\partial}{\partial t}y^{\prime}(t,s,s^{\prime})=-\sum\limits _{j=1}^{N}y_{j}(t,s)Y^{\prime}(t,s^{\prime}),\label{eq:ptyp}
\end{equation}
where $X_{i}(t)=\int_{0}^{t}K_{1}(t,s)x_{i}(t,s)ds,$ $X^{\prime}(t,s^{\prime})=\int_{0}^{t}K_{1}(t,s)x^{\prime}(t,s,s^{\prime})ds,$
$Y_{i}(t)=\int_{0}^{t}K_{2}(t,s)y_{i}(t,s)ds,$ and $Y^{\prime}(t,s^{\prime})=\int_{0}^{t}K_{2}(t,s)y^{\prime}(t,s,s^{\prime})ds,$
with the initial conditions $x_{i}(t,t)=1$, $y_{i}(t,t)=1$, $x^{\prime}(t,t,s^{\prime})=y^{\prime}(t,t,s^{\prime})=0,$
$x^{\prime}(t,s,t)=-{\textstyle \sum\nolimits _{j=1}^{N}}x_{j}(t,s),$
and $y^{\prime}(t,s,t)=\sum\nolimits _{j=1}^{N}y_{j}(t,s).$ With
the $O$ operators and the coefficients in Eqs.~(\ref{FTO1}-\ref{eq:ptyp}),
one can numerically simulate the evolution of the cavity array in
finite temperature case by using Eq.~(\ref{eq:QSDFT}). Certainly,
a master equation can be also obtained by following the method in
Refs.~\cite{Chen2014PRA,Strunz2004PRA}.

The numerical results (not presented) show the temperature effect can enhance or weaken the transfer. However, in this paper, we focus on some distinctive difference between non-Markovian and Markovian environments. Instead of ``enhance or weaken'', we pay more attention to ``can or cannot''. The detailed influence of the temperature can be investigated elsewhere in the future.

\begin{backmatter}
\bmsection{Funding}

National Natural Science Foundation of China under Grant Nos. 62471143 and 11874114

Natural Science Funds for Distinguished Young Scholar of Fujian Province under Grant
2020J06011

Natural Science Foundation of Fujian Province under Grant
No. 2022J01548

Project from Fuzhou University under Grant No. JG202001-2
and Grant No. GXRC-21014.

\bmsection{Acknowledgments}
This work was supported by the National Natural Science Foundation
of China under Grant Nos. 62471143 and 11874114, the Natural Science
Funds for Distinguished Young Scholar of Fujian Province under Grant
2020J06011, Natural Science Foundation of Fujian Province under Grant
No. 2022J01548, Project from Fuzhou University under Grant No. JG202001-2
and Grant No. GXRC-21014.

\bmsection{Disclosures}
The authors declare no conflicts of interest.

\end{backmatter}

\bibliography{cat.bib}

\end{document}